\documentstyle[11pt,epsf]{article}

\newcommand{\be}{\begin{equation}}
\newcommand{\ee}{\end{equation}}
\newcommand{\nn}{\nonumber}
\newcommand{\beba}{\begin{equation}\begin{array}{lcl}}
\newcommand{\eaee}{\end{array}\end{equation}}
\newcommand{\bea}{\begin{eqnarray}}
\newcommand{\eea}{\end{eqnarray}}
\newcommand{\ba}{\begin{array}}
\newcommand{\ea}{\end{array}}

\newcommand{\ns}{\normalsize}
\newcommand{\refs}[1]{(\ref{#1})}
\def\bal{{\mbox{\boldmath $\alpha$}}}

\def\bq{{\bf q}}
\def\bd{{\bf d}}
\def\bk{{\bf k}}
\def\bc{{\bf c}}
\def\bw{{\bf w}}
\def\bH{{\bf H}}
\def\bp{{\bf p}}
\def\bk{{\bf k}}
\def\bx{{\bf x}}

\def\a{\alpha}
\def\b{\beta}
\def\g{\gamma}

\def\d{\delta}
\def\e{\epsilon}

\def\f{\phi}

\def\l{\lambda}
\def\m{\mu}
\def\n{\nu}

\def\r{\rho}
\def\s{\sigma}

\def\t{\tau}

\def\D{\Delta}
\def\F{\Phi}

\def\L{\Lambda}
\def\O{\Omega}

\def\ght{\hat{g}\kern-0.6em \widetilde{\raisebox{-0.12em}{\phantom{X}}}}
\def\bht{\hat{b}\kern-0.6em \widetilde{\raisebox{0.15em}{\phantom{X}}}}

\def\cL{{\cal L}}
\def\cM{{\cal M}}

\begin{document}

\begin{titlepage}
\title{\hfill{\ns UPR-792T\\}
       \hfill{\ns hep-th/9802041\\[.1cm]}
       \hfill{\ns February 1998}\\[.8cm]
       {\large\bf The Cosmology of M-Theory and Type II Superstrings}}
\author{Andr\'e
        Lukas$^1$\setcounter{footnote}{0}\thanks{Supported by Deutsche
        Forschungsgemeinschaft (DFG).}~~,
        Burt A.~Ovrut$^1$\thanks
        {Based on invited talks at the Europhysics Conference on High Energy 
        Physics, August 19-26, 1997, Jerusalem, Israel and at the XXXIIIrd
        Rencontres de Moriond, Fundamental Parameters in Cosmology, Les Arcs,
        Savoie, France, January 17-24, 1998.}~~
        and Daniel Waldram$^2$\\[0.5cm]
        {\ns $^1$Department of Physics, University of Pennsylvania} \\
        {\ns Philadelphia, PA 19104--6396, USA}\\[0.3cm]
        {\ns $^2$Department of Physics}\\
        {\ns Joseph Henry Laboratories, Princeton University}\\
        {\ns Princeton, NJ 08544, USA}}
        
\date{}
\maketitle

\begin{abstract} 
\thispagestyle{empty}

We review cosmological solutions of type II superstrings and M-theory,
emphasizing the role of non-vanishing Ramond form backgrounds.
Compactifications on flat and, more generally, maximally symmetric spatial
subspaces are presented. We give a physical discussion of both inflating and
subluminally expanding cosmological solutions of such theories and explore
their singularity structure. An explicit example, in the context of the
type IIA  superstring, is constructed. We then analyze
compactifications of M-theory on Ricci flat manifolds. The external part
of U--duality and its relation to cosmological solutions is studied in
the low energy theory. In particular, we investigate the behaviour of
important cosmological properties, such as the Hubble parameters and the
transition time between two asymptotic regions, under U-duality
transformations. Motivated by Horava-Witten theory, we present an
explicit example of manifestly U-duality covariant M--theory cosmology
in a five-dimensional model resulting from compactification on a Calabi-Yau
three-fold.

\end{abstract}

\thispagestyle{empty}
\end{titlepage}


\section{Introduction}

An important constraint on string theory or any generalization of
string theory, such as M--theory, is that it should be compatible with
the standard model of early universe cosmology. 
In the past, most focus has been on the weakly coupled 
heterotic string as the best
model of low-energy particle physics. However, with the
discovery of string dualities \cite{string_dual} and the
existence of D--brane states \cite{D-brane}, the
nature of string theory has changed dramatically. Strong-weak coupling
duality symmetries connect each of the five consistent supersymmetric
string theories together with eleven-dimensional supergravity
\cite{duff_rep}. As a consequence, type II and eleven-dimensional
supergravities may now be directly relevant to low-energy particle physics
and cosmology \cite{banks_dine}. Both theories contain form fields,
namely a three form in M-theory and Ramond-Ramond (RR) forms of
various degrees in type II theories. Given this change of perspective, it
clearly becomes important to study the cosmological solutions of type
II superstrings and M-theory with non-trivial form fields
excited. The present paper discusses a general framework for
analyzing cosmologies of such theories. This framework
has been presented in detail in~\cite{letter,paper,dil,ucos}.

Various aspects of M--theory cosmology have been studied over the past
two years such as cosmological solutions with nontrivial form
fields~\cite{letter,ps,kal,lu1,paper,lw,lu2,kko}, the possibility of
singularity free solutions~\cite{paper,lw,ra,kfs,ko,kko} and moduli and
dilaton stabilization~\cite{dil}.
Applications of T--duality and S--duality to string cosmology have
recently been discussed in the ref.~\cite{ax_dil,li,cew1,clw0,ven,ba}.
In particular, the $SL(2)$ symmetry of type IIB has been used to
generate cosmological solutions with Ramond--Ramond fields~\cite{ps,sen,clw}.
The S--duality properties of cosmological quantities have been analyzed
in~\cite{cew1,clw0,clw}. Cosmological solutions with Ramond forms
obtained from black hole solutions have been constructed
in~\cite{be_fo,ps,lw}.

In reference~\cite{letter,paper}, we considered compactifications on
maximally symmetric subspaces. The first part of the present paper is a
review of this formalism. For such compactifications,
our general result is that the
moduli-space potential which arises when non-trivial form-fields are
excited, significantly effects the structure of the solution. 
When all the subspaces are flat, the potential is operative over a
particular, finite part of the evolution. 
At the extremes we return to the simple Kaluza-Klein Kasner-type~\cite{kasner}
solutions with some subspaces expanding, some contracting. Consequently,
there is always either an initial or a final curvature singularity. Thus the
effect of the form fields is to interpolate between two different
Kaluza-Klein solutions. In general, for solutions with an initial 
singularity, one finds
that the rate of expansion is always sub-luminal and so there is no
inflation. On the other hand, those solutions with a final
singularity, just as in the pre-big-bang models \cite{pre_BB}, may exhibit
superinflation but are unphysical as stands because the inflation ends
in a curvature singularity. 
The solution can be very different, however, when we allow for 
subspaces of non-vanishing constant curvature. 
The effect of the curvature is to introduce new terms into
the moduli space potential. This
can significantly change the singularity structure, in some cases giving
solutions where the curvature always remains finite. This suggest the
interesting possibility that there may exist inflating solutions which
are not forced to end in a singularity.

In reference~\cite{ucos}, we analyzed U-duality covariant compactifications
on Ricci flat manifolds in general and Calabi-Yau manifolds in particular.
In the second part of this paper we discuss the results of~\cite{ucos}.
Specifically , we will present a manifestly U--duality covariant
formulation of M--theory cosmology. Since U--duality rotates metric degrees
of freedom and degrees of freedom from the 3--form of 11-dimensional
supergravity into each other, we will be naturally dealing with cosmologies
which have nontrivial Ramond--Ramond fields. 
The most interesting part of duality transformations in cosmology is the one
which acts non--trivially on the space--time metric and it is this part
on which we will concentrate. Therefore, we are going to reduce 11--dimensional
supergravity on a Ricci--flat manifold to $D$ dimensions, thereby keeping the
breathing mode of the internal space as the only modulus~\cite{maeda} and
focusing on the $D$--dimensional ``external'' part of U--duality. As a general
rule, this part of the U--duality group acting on cosmological solutions in
$D$ space--time dimensions is the same as the U--duality group of
$12-D$--dimensional maximal supergravity. As an explicit example, we will
study the case $D=5$, corresponding to the U--duality group
$G=SL(5)$. This example is
motivated by the Horava--Witten construction of M--theory on
$S^1/Z_2$ ~\cite{hw} which represents the effective theory
of strongly coupled heterotic string theory. This theory turns out to be
effectively 5--dimensional for phenomenological values of the coupling
constants in some intermediate energy range ~\cite{w}. 

As a result, we find large classes of cosmological solutions which are
mapped into each other by U--duality transformations. The most characteristic
properties of these solutions are very similar to the ones found for
maximally symmetric subspaces. They usually have two different branches,
one with a future and the other with a past curvature singularity.
Each branch consists of asymptotic Kaluza--Klein regions which
evolve into each other under the influence of the form field. We find
that U--duality relates to this structure in an interesting way. In
the asymptotic Kaluza--Klein regions the basic expansion properties such
as Hubble parameters turn out to be U--duality invariants. A related
phenomenon, in the context of S--duality, has been observed in
ref.~\cite{cew1,clw0,clw}. U--duality transformations, however, do change
the transition time between asymptotic regions and influence the details
of the transition. Using the knowledge on how U--duality acts explicitly
on the cosmological solutions we, moreover, show that a U--duality version of
the pre--big--bang scenario~\cite{pbb} can be realized within our setting.


\section{Solutions with Maximally Symmetric Subspaces -- General
         Framework}

In this section, we present a general framework for finding
cosmological solutions with non-trivial form fields compactified
on subspaces of constant curvature. 
The starting point for our investigation is the following effective action 
\be
 \bar{S} = \int d^Dx\sqrt{\bar{g}}\left[e^{-2\f}(\bar{R}+4(\partial\f )^2
           -\frac{1}{12}H^2)-\sum_r \frac{1}{2(\d_r+1)!}F_r^2-\L \right]
 \label{string_action}
\ee
with the $D$--dimensional string frame metric $\bar{g}_{MN}$, the dilaton
$\f$, the NS 2--form $H$ and a number of RR $\d_r$--forms $F_r=dA_r$. We
also allow for a cosmological constant $\L$ which appears in the massive
extension of IIA supergravity~\cite{romans}. Assuming this origin, it
is restricted to be positive, $\L >0$. 

The above action can account for a wide range of cosmological solutions
in type II theories (where we usually have $D=10$ in mind) and, if the
dilaton is set to zero, in $D=11$ supergravity. For simplicity we have
kept only the kinetic terms for the form fields. In general, in both
type II theories and eleven dimensional supergravity there are
additional terms involving the coupling of form fields. We shall
assume throughout this paper that for the configurations with which we
are concerned, these terms do not contribute to the equations of
motion. 
\vspace{0.4cm}

In order to give a physical description of our solutions, we prefer to work
in the canonical Einstein frame metric $g_{MN}$ which is related to
the string frame metric by a conformal rescaling $g_{MN}=\exp (-4/(D-2)\f
)\bar{g}_{MN}$. The corresponding action reads
\be
 S=\int d^Dx\;\sqrt{-g}\left[ R-\frac{4}{D-2}(\partial\f )^2-\sum_r
    \frac{1}{2(\d_r+1)!}e^{-a(\d_r)\f}F_r^2-\L e^{-a_\L \f}\right]
    \label{action}
\ee
where the NS field $H$ has been included in the sum over $F_r$. It is
distinguished from the other forms by the dilaton couplings $a(\d_r)$ given
by
\be
 a(\d_r) = \left\{ \ba{cll} \frac{8}{D-2}&{\rm NS}&{\rm 2-form}\\
                      \frac{4\d_r -2(D-2)}{D-2}&{\rm RR}&\d_r{\rm -form}
                \ea\right.\; . \label{p_rr}
\ee
The coupling $a_\L$ for the cosmological constant
\be
 a_\L = -\frac{2D}{D-2}  \label{p_lambda}
\ee
equals the negative of the one for a RR $(D-1)$--form. 

\vspace{0.4cm}

The type of solutions we consider are characterized by a space split
into $n$ flat subspaces, each of them characterized
by a scale factor $\a_i$. We concentrate on {\em flat} subspaces in this
paper.
The effect of
{\em curved} maximally symmetric subspaces can easily be incorporated
into the formalism. In the flat case, the corresponding metric is given by
\be
 ds^2 = -N^2(\t )d\t^2+\sum_{i=0}^{n-1}e^{2\a_i(\t )}d{\bf x}_i^2
 \label{metric}
\ee
where $d{\bf x}_i^2$ is the measure of a $d_i$-dimensional flat maximally
symmetric subspace and $\sum_{i=0}^{n-1}d_i=D-1$. For solutions with
this structure, the
dilaton should depend on time only, $\f =\f (\t )$. This Kaluza--Klein-type
Ansatz is about the simplest allowing for the cosmologically key properties
of homogeneity and isotropy as well as for an ``external'' and ``internal''
space. 

For a ten dimensional theory, the simplest choice is to split up the
space into two subspaces ($n=2$) 
with $d_0=3$ and $d_1=6$. Then the $d_0=3$ part could be interpreted as
the spatial part of ``our'' 4--dimensional  space--time with an evolution
described by the scale factor $\a_0$. The other six dimensions would form an
internal space with a modulus $\a_1$. Clearly, one is free to split these
six dimensions even further or even allow for a further split of the
3--dimensional space.

The choice of subspaces is important in fixing the possible
structure of the antisymmetric tensors fields~\cite{freund_rubin,freund}. 
The symmetry of the above metric allows for two different Ans\"atze
for the form fields which we call ``elementary'' and ``solitonic'' in
analogy to the two types of $p$--brane solutions. They are
characterized by the following nonvanishing components of the field
strength. 

\begin{itemize}
 \item elementary~: if $\sum_i d_i=\d_r$ for some of the spatial subspaces $i$
                    we may set
 \be
  (F_r)_{0\m_1...\m_{\d_r}} = A_r(\a )\, f_r'(\t )\,\e_{\m_1...\m_{\d_r}}\; ,
  \quad A_r(\a ) = e^{ -2\sum_i d_i\a_i } \label{elementary}
 \ee
 where $\m_1...\m_{\d_r}$ refer to the coordinates of these subspaces,
 $f_r(\t )$ is an arbitrary function to be fixed by the form field
 equation of motion, and the prime denotes the derivative with
 respect to $\t$. With raised indices, the symbol
 $\e^{\m_1...\m_{\d_r}}$ takes the values 0 or 1 and is completely 
 antisymmetric on all $\d_r$ indices. Note 
 that the sum over $i$ in the exponent runs only over
 those subspaces which are spanned by the form.
\end{itemize}
An elementary form can therefore extend over one or more of the
subspaces only if its degree matches the total dimension of these spaces.
Consider for example the RR three-form of type IIA supergravity. If the space
is split up as $(d_0,d_1)=(3,6)$ it fits into the 3--dimensional subspace
and the above general Ansatz specializes to
$F_{0\m_1\m_2\m_3}=\exp (-6\a_0)f'(\t )\e_{\m_1\m_2\m_3}$ where
$\m_1,\m_2,\m_3$ refer to the coordinates of this subspace. 

\begin{itemize} 
 \item solitonic~: if $\sum_i d_i=\d_r+1$ for some of the spatial subspaces $i$
                   we may allow for
 \be
  (F_r)_{\m_1...\m_{\d_r+1}} = B_r(\a )\; w_r\; \e_{\m_1...\m_{\d_r+1}}\;
 ,\quad
  B_r(\a ) = e^{ -2\sum_i d_i\a_i } \label{solitonic}
 \ee
 where $\m_1...\m_{\d_r+1}$ refer to the coordinates of these subspaces and
 $w_r$ is an arbitrary constant. As for the elementary Ansatz, the sum
 over $i$ in the exponent runs over the subspaces spanned by the
 form. It is easy to check that this Ansatz, already solves the form
 equation of motion. 
\end{itemize}
Note that in contrast to the elementary Ansatz, the solitonic field strength
does not have a time index. Therefore the matching condition for the
dimensions differs. Given, for example, a split $(d_0,d_1)=(3,6)$ one has to
use a  2 form instead of a three-form to fit into the 3-dimensional subspace.
The above Ansatz then reads
$F_{\m_1\m_2\m_3}=\exp (-6\a_0)w\e_{\m_1\m_2\m_3}$.

\vspace{0.4cm}

Having specified the form of our Ansatz, we now look to solve the
equations of motion derived from the
action~\refs{action}. However, it is in fact easy to show that,
under a very mild restriction, the resulting equations of 
motion can be obtained from a 
reduced Lagrangian which depends only on $\a_i$, $\f$, $N$ and
$f_r$, each as functions of $\t$. The Lagrangian is given by
\be
 {\cal L} = E\left[\sum_{i=0}^{n-1}d_i\a_i'^2-
            \sum_{i,j=0}^{n-1}d_id_j\a_i'\a_j'
            +\frac{4}{D-2}\f '^2+V_e-N^2V_s\right]
 \label{lagrangian}
\ee
with
\bea
 V_e &=& \frac{1}{2}\sum_{r}A_r(\a )e^{-a(\d_r)\f}{f_r'}^2 \nn\\
 V_s &=& \frac{1}{2}\sum_{r}B_r(\a )w_r^2e^{-a(\d_r)\f}+\L
         e^{-a_\L\f} \label{Ves} \\
 E &=& \frac{1}{N}e^{\sum_{i=0}^{n-1}d_i\a_i }\; . \nn
\eea
In the definitions of the potentials $V_e$ and $V_s$, the sum over $r$ is
understood to run over all the elementary and solitonic configurations
which have been chosen according to the given rules. 
The equations of motion for the functions $f_r$ to be derived from
eq.~\refs{lagrangian} read
\be
 \frac{d}{d\t}\left( EA_re^{-a(\d_r)\f}f_r'\right) = 0 \; .
\ee
The first integrals are
\be
 f_r' = v_rE^{-1}A_r^{-1}e^{a(\d_r)\f} \label{fr}
\ee 
where $v_r$ are integration constants. Equation~\refs{fr} can be used to
eliminate $f_r'$ from the elementary potential $V_e$. This then
reduces the problem to solving the remaining equations of motion now
given purely in terms of $\a_i$, $\f$ and $N$. 

In fact, we find that the remaining equations can also be derived from
a simple reduced Lagrangian. First introduce the notation 
$\bal=(\a_I)=(\a_i,\f )$ for a general
point in the moduli space. We also define a particular metric on the
moduli space $G_{IJ}$ by 
\bea
 G_{ij}&=&2(d_i\d_{ij}-d_id_j)\nn \\
 G_{in}&=&G_{ni}=0 \label{G}\\
 G_{nn}&=&\frac{8}{D-2}\; .\nn 
\eea
The equations of motion for $\bal$ and $N$ following from
eq.~\refs{lagrangian} then take the simple form
\bea
 \frac{d}{d\t}\left( EG\bal '\right)+E^{-1}\frac{\partial U}{\partial\bal}
  &=&0 \label{al_eom} \\ 
 \frac{1}{2}E{\bal '}^TG\bal '+E^{-1}U &=& 0\; . \label{N_eom}
\eea
The quantity $U$ is given by
\be
 U=e^{2\sum_{i=0}^{n-1}d_i\a_i}\left( \frac{1}{N^2}V_e+V_s\right)
 \label{pot_def}
\ee
where $f_r'$ in $V_e$ has been replaced using eq.~\refs{fr}. Clearly,
these equations of motion can be derived from the simple Lagrangian
(it is convenient to make the change of variables from $N$ to $E$),
\be
 {\cal L} = \frac{1}{2}E{\bal '}^TG\bal '-E^{-1}U \label{dan_lag}
\ee
The first term is
kinetic, while $U$ defines a potential in the moduli space. Further, 
$E$ is the metric on the particle worldline. 

It is useful to rewrite the potential $U$ in a more systematic
way as
\be
 U = \frac{1}{2}\sum_{r=1}^{m}u_r^2\exp (\bq_r \cdot\bal ) \label{U}
\ee
where the sum runs over all elementary and solitonic configurations as well
as a possible cosmological constant term. The constants $u_r$ represent
the integration constants $v_r$ in eq.~\refs{fr} for elementary forms,
the constants $w_r$ in the Ansatz~\refs{solitonic} for solitonic
forms or a cosmological constant. The type of each term is specified
by the vectors $\bq_r$ which can be read off from eqs.~\refs{elementary},
\refs{solitonic}, \refs{Ves} and the definition~\refs{pot_def}. For an
{\em elementary} $\d$--form they are given by
\be
 \bq^{\rm (el)} = (2\e_id_i,a(\d ))\; ,\quad \e_i=0,1\; ,
 \quad \d =\sum_{i=0}^{n-1}\e_id_i \label{q_el}
\ee
with $\e_i=1$ if the form is nonvanishing in the subspace $i$ and
$\e_i=0$ otherwise. For type II theories the dilaton couplings $a(\d )$ are
given in eq.~\refs{p_rr} and~\refs{p_lambda}. To account for the $D=11$
case (or constant dilaton solutions) we may just set $a(\d )=0$.

Let us give an example at this point. An elementary IIA RR 3 form, put
into the first subspace of a $D=10$ space split with $(d_0,d_1)=(3,6)$, is
characterized by a vector $\bq = (6,0,-1/2)$. It generates a
potential term in~\refs{U} which depends on $\a_0$ and the dilaton but not
on $\a_1$. More generally, a potential term describing
the effect of an elementary form depends only on those scale factors
which correspond to subspaces spanned by the form. Since the
entries $q_i$ of $\bq$ are always positive, the potential tends to
drive the scale factors for these subspaces to smaller values. Therefore
these subspaces tend to be contracting or at least less generically
expanding than others.

The situation for a {\em solitonic} $\d$--form is in some sense complementary.
It is specified by a vector
\be 
 \bq^{\rm (sol)} = (2\tilde{\e}_id_i,-a(\d ))\; ,\quad
  \tilde{\e}_i\equiv 1-\e_i
 =0,1\; ,\quad \tilde{\d}\equiv D-2-\d =\sum_{i=0}^{n-1}\tilde{\e}_id_i
 \label{q_sol}
\ee
with $\tilde{\e}_i=1$ if the form vanishes in the subspace $i$ and
$\tilde{\e}_i=0$ otherwise. For example, a solitonic IIB RR 2 form in the
first subspace of a space split as $(d_0,d_1)=(3,6)$ is specified
by $\bq = (0,12,1)$. It generates a potential term in~\refs{U} which depends
on $\a_1$ and the dilaton but not on $\a_0$. More generally, in contrast to
the elementary case, the potential term now depends on those scale factors
corresponding to subspaces {\em not} spanned by the form.

Finally, a cosmological constant is characterized by
\be
 \bq^{(\L )} = \left( 2d_i,\frac{2D}{D-2}\right)\label{q_lambda}\; .
\ee
Note that for all these vectors $\sum_{i=0}^{n-1}q_i>0$, a fact which
we will use later on. The moduli space metric allows us to define a
natural scalar product on the space of vectors $\bq$
\be
 <\bq_1 ,\bq_2 > = \bq_1^TG^{-1}\bq_2 \label{s_prod}
\ee
with the inverse of $G$ given by
\bea
 (G^{-1})_{ij} &=& -\frac{1}{2(D-2)}+\frac{1}{2d_i}\d_{ij} \nn \\
 (G^{-1})_{in} &=& (G^{-1})_{ni} = 0 \label{Gin}\\
 (G^{-1})_{nn} &=& \frac{D-2}{8} \; . \nn
\eea
Since the metric $G$ has Minkowskian signature, we can 
distinguish between space- and time-like vectors $\bq$. As we will see,
the structure of the solutions depends crucially on this distinction.

\vspace{0.4cm}

Generically, the models specified by the eqs.~\refs{al_eom}, \refs{N_eom}
and~\refs{U} cannot be solved. A complete solution, however, can be found
if the potential $U$ consists of one exponential term only or if contact
with Toda theory can be made. Here, we will discuss only the first of these
two possibilities. For examples related to Toda theory see
ref.~\cite{letter,paper}.


\section{Solutions with One Potential Term} 

In this section, we will analyze models with just one form turned
on (or a non-zero cosmological constant). The form may be elementary or
fundamental and there may be any number of subspaces. All of these
cases correspond to a potential 
\be
 U = \frac{1}{2}u^2\exp (\bq \cdot\bal ) \label{U1}
\ee
where $u^2>0$. We will start by giving the general form of the
solution and then give a simple example in section 3.2.

One way of solving the equations of motion for a potential~\refs{U1} is
to use the gauge freedom in the definition of the time coordinate. We
can always choose a gauge such that 
\be
 N=\exp ((\bd -\bq )\cdot\bal ) \label{gauge}
\ee
where we introduce a vector giving the subspace dimensions $\bd =
(d_i,0)$. This implies $E = \exp (\bq \cdot\bal )$ and the following
set of equations for $\bal$ 
\bea
 \frac{d}{d\t}(GE\bal ')+\frac{1}{2}u^2\bq &=& 0 \nn \\
 \frac{E}{2}{\bal^T}'G\bal ' +\frac{1}{2}u^2 &=& 0\; \label{eom_one}.
\eea
In this form they can be integrated immediately, leading to the general
solution
\be
 \bal = \bc\ln |\t_1 -\t |+\bw\ln\left(\frac{s\t}{\t_1-\t}\right)+\bk
 \label{sol1}
\ee
where
\be
 \bc = \frac{2G^{-1}\bq}{<\bq ,\bq >}\; . \label{c}
\ee
The sign $s=\pm 1$ is determined by
$s = {\rm sign}(<\bq ,\bq >)$~\footnote{Here we disregarded the somewhat
marginal possibility that $\bq$ is a null vector, i.~e. $<\bq ,\bq >=0$.}
and $\bw$, $\bk$ are integration constants subject to the constraints
\bea
 \bq \cdot\bw &=& 1 \nn \\
 \bw^TG\bw &=& 0 \label{cons1}\\
 \bq \cdot \bk &=& \ln\left(\frac{u^2|<\bq ,\bq >|}{4}\right)\; . \nn
\eea
$\t_1$ is a free parameter which we can take to be positive. 
The range of $\t$ should be chosen to ensure
a positive argument of the second logarithm in eq.~\refs{sol1}. This depends
on the sign of $<\bq ,\bq >$ and we have the two cases 
\be
 \ba{ccc}
  0<\t <\t_1&{\rm for}&<\bq ,\bq >\; > 0\\
  \t <0\;{\rm or}\; \t>\t_1&{\rm for}&<\bq ,\bq >\; < 0
 \ea\; . \label{ranges}
\ee
Which of these cases is actually realized in type II models? Using
the vectors $\bq$ given in the end of section 2, we find for
a solitonic or elementary $\d$ form (or a cosmological constant which
is similar to a RR $(D-1)$--form)
\be
 <\bq ,\bq > = \frac{D-2}{8}a(\d )^2+\frac{2}{D-2}\d\tilde{\d}=\left\{
               \ba{cll} 4&&{\rm NS} \\
                        \frac{D-2}{2}&&{\rm R}
               \ea \right. \label{qq}
\ee
which is always positive. Also the $D=11$ 3 form leads to
a positive result, as can be seen from the above formula by setting $a(\d )=0$.
We conclude that, in the present context, we are dealing with spacelike
vectors $\bq$ only, and we have $0<\t<\t_1$.


\vspace{0.4cm}

So far, our solutions have been expressed in terms of the time parameter
$\t$ which is defined by the gauge choice~\refs{gauge} for $N(\t )$.
For a discussion of the cosmological properties of our models,
however, we should reexpress them in terms of the comoving time $t$,
that is in the gauge where the $N=1$. This can
be done by integrating the defining relation $dt=N(\t )d\t$. 
The explicit expression for $N(\t )$ is given 
by inserting the solution~\refs{sol1} into the gauge fixing
equation~\refs{gauge} for $N(\t )$, which gives 
\be
 N = \exp ((\bd -\bq )\cdot\bk )|\t_1-\t |^{-x+\D -1}||\t |^{x-1} \label{N}\\
\ee
with
\bea
 x &=& \bd \cdot\bw \label{x}\\
 \D &=& \bd \cdot\bc = 2\frac{<\bd ,\bq >}{<\bq ,\bq >} \; . \label{Delta}
\eea
Depending on the values of
$x$ and $\D$, the gauge parameter $N$ may have singularities at $\t =0$ and
$\t =\t_1$. This determines the allowed range in the comoving time $t$ as we
will discuss in detail in the next subsection.

Another quantity which is of importance in discussing the physical
content of our solutions is the scalar curvature $R$. For the
solution~\refs{sol1} it is given by
\be
R \sim |\t_1 -\t |^{2(x-\D )}|\t |^{-2x}P_2(\bw ,\bal ,\t ) \label{R}
\ee
where $P_2$ is a second order polynomial in $\t$ which we will not need
explicitly. The first two factors in this
equation indicate potential singularities at $\t = 0$ and $\t =\t_1$,
depending on $x$ and $\D$ as in the case of the
gauge parameter $N$. However, in contrast to singularities in $N$, such
singularities are true coordinate independent curvature singularities.
They will be further discussed in the next subsection. 


\subsection{Cosmology of Solutions with Spacelike $\bq$--Vectors}

As already mentioned, the case of spacelike $\bq$-vectors 
is the most important in our context since
all vectors $\bq$ appearing within the $D=10$ type II theories and 
M-theory are spacelike. 
In this section, we will discuss the cosmological properties which can be
extracted from these models in general. A concrete illustrating example
will be given in the next subsection.

Recall that the singularity structure 
of the solution~\refs{sol1} is determined by the
quantities $x$ and $\D$ defined in eq.~\refs{x} and~\refs{Delta}. What values
are actually allowed for these quantities?
{}From $<\bd ,\bq >=-\sum_{j=0}^{n-1}q_j/2(D-2)$ and $\sum_{j=0}^{n-1}q_j>0$
it follows from eq.~\refs{Delta} that $\D <0$. The parameter 
$x$, which unlike $\D$ depends on
the parameters of the solution, turns out to be either $x<\D$
or $x>0$ in all specific examples we considered. This divides the set
of initial conditions into two disconnected subsets corresponding to
two classes of solutions with different properties. 

We begin our discussion of these properties by analyzing the allowed ranges
in comoving time $t$. Recall from eq.~\refs{ranges} that the time parameter
$\t$ is always in the range $0<\t <\t_1$ since we have $<\bq ,\bq >\;
>0$. The singularity structure of the gauge parameter $N$ in
eq.~\refs{N} then shows that this range is mapped to the following
ranges in $t$ 
\be
 \t\rightarrow t\in\left\{\ba{clll}
       \left[ -\infty ,t_1\right]&{\rm for}\; x<\D<0\; ,&(-)\;{\rm branch} \\
       \left[ t_0,+\infty\right]&{\rm for}\; x>0>\D\; ,&(+)\;{\rm branch}
       \ea\right.\; .
\ee
Here $t_0$ and $t_1$ are two finite unrelated values that appear as
integration constants from integrating $dt =N(\t )d\t$. Thus we have found
two disconnected branches corresponding to asymptotically positive and
negative time ranges.

Let us next discuss the scalar curvature in each branch. We start with
the $(-)$ branch. As inspection of eq.~\refs{R} shows, the curvature
vanishes as $t\rightarrow -\infty$ ($\t\simeq 0$) since $x<\D <0$. With
increasing time $R$ grows and, finally, the system runs into the curvature
singularity at $t=t_1$ ($\t =\t_1$) since the power $2(x-\D )$ of the
first term in eq.~\refs{R} is negative. Therefore, classically the
$(-)$ solution cannot be continued beyond this point.

In the $(+)$ branch the situation is similar but reversed in time. At
$t=t_0$ ($\t =0$) we find a singularity since $x>0$ in this branch.
The solution cannot be extended into the past. As $t\rightarrow\infty$
($\t\simeq\t_1$) the curvature behaves smoothly and approaches zero.

Though generically correct, the above argument has a loophole.
For very specific values of the initial parameters $\bw$, the polynomial
$P_2$ is proportional to $\t$ or $\t_1-\t$ so that it can cancel against
one of the first two factors in eq.~\refs{R} which cause the singularity.
If $|x|$ is sufficiently small, the singularity may disappear completely.
For the $(-)$ branch this is realized if $w_n=c_n$ and $x\geq -1/2$. For
the $(+)$ branch it occurs if $w_n=0$ and $x\leq 1/2$.
This phenomenon is quite similar to what happens in the curvature
singularity free WZW model of ref.~\cite{kou_lust}

\vspace{0.4cm}

So far, we have  considered quantities which provide information
about the behaviour of the total $D$ dimensional space only. Let us now turn
to the individual subspaces of dimension $d_i$. To analyze their
behaviour, we should calculate their respective Hubble parameters $H_i$ in
terms of the comoving time. In fact, it is possible to explicitly
express the comoving time $t(\t )$ in terms of hypergeometric functions.
It is, however, more instructive to look at the asymptotic regions $\t\simeq 0$
(corresponding to $t\rightarrow -\infty$ for the $(-)$ branch and
$t\simeq t_0$ for the $(+)$ branch) and $\t\simeq\t_1$
(corresponding to $t\simeq t_1$ for the $(-)$ branch and
$t\rightarrow\infty$ for the $(+)$ branch). In these regions the Hubble
parameters can be written as~\footnote{The dot denotes the derivative with
respect to the comoving time $t$.}
\be
 \bH \equiv \dot{\bal} = \frac{\bp}{t-t_s} \label{hubble}
\ee
with the constant expansion coefficients $\bp$ satisfying
\be
 \bp G\bp = 0\; ,\quad \bd \cdot\bp = 1\; .\label{kk}
\ee
The time shift $t_s$ depends on the asymptotic region and branch under
consideration.
The sign of $t-t_s$, however, is always well defined~: It is negative in
the $(-)$ branch and positive in the $(+)$ branch. If we combine the
two equations~\refs{kk} and use the explicit form of the metric
$G$ in~\refs{G} we find
\be
 \frac{4}{D-2}p_\f^2+\sum_{i=0}^{n-1}d_ip_i^2 = 1\; .\label{p_bound}
\ee
The explicit expressions for $\bp$ in terms of the integration constants
are 
\be
 \bp = \left\{\ba{cll} \frac{\bw}{x}&{\rm at}&\t\simeq 0 \\
                       \frac{\bw -\bc}{x-\D}&{\rm at}&\t\simeq\t_1
       \ea\right. \; . \label{p_expr}
\ee
They have been calculated using the general solution~\refs{sol1} and
the asymptotic limits of $N(\t )$ to be read off from eq.~\refs{N}.
The behavior of the Hubble parameters~\refs{hubble} along with eq.~\refs{kk}
indicates that the solutions behave asymptotically like those of pure
Kaluza--Klein theory with a dilaton. This can be seen by a comparison with
the solutions of ref.~\cite{mueller}. Therefore, one expects that the
potential $U$ provided by the form is effectively turned off in these
limits. In fact, inserting the general solution~\refs{sol1} into the
potential~\refs{U1}, we find
$U\sim (\t_1-\t )\t$ which implies that $U$ is effectively zero near
$\t\simeq 0$ and $\t\simeq \t_1$. The effect of the form is therefore
to generate a mapping $\bp (\t\simeq 0)\rightarrow\bp (\t\simeq\t_1)$
between two pure Kaluza--Klein states.

What do the above results mean for the evolution of the subspaces? We
consider the $(+)$ branch first. Remember that $t-t_s>0$ in this branch
so that from eq.~\refs{hubble} a positive $p_i$ results in expansion
and a negative $p_i$ in contraction. Moreover, the equation $\bd\cdot\bp = 1$
shows that at least one of the $p_i$ has to be positive. Consequently, at
least one of the subspaces has to be expanding. From eq.~\refs{p_bound} we
conclude that $|p_i|<1$ always. The expansion is therefore subluminal.
This behaviour is similar to a radiation or matter
dominated universe corresponding to $p_i=1/2$ and $p_i=2/3$, respectively.

The situation is completely different in the $(-)$ branch. Since
$t-t_s<0$, a positive $p_i$ results in contraction and a negative $p_i$
in expansion. Now we conclude from $\bd\cdot\bp =1$ that at least one
subspace must be contracting. Since we are in the negative time range,
eq.~\refs{hubble} shows that expansion ($H_i>0$) goes along with an
increasing $H_i$, that is, a shrinking horizon size. Such a behaviour
is also called superinflation since scales are stretched across the
horizon even more rapidly than in ``ordinary'' inflation where the horizon
size is approximately constant. 

Our solutions allow various patterns of expanding and contracting spatial
subspaces. The details of the evolution depend on the partition $\{ d_i\}$,
the form and the subspaces it occupies and the initial conditions.
Examples with $3$ expanding and $6$ contracting spatial dimensions
as $t\rightarrow\infty$ exist, as were given in
ref.~\cite{letter}. The effect of
the form can be quite dramatic. For example, it can reverse expansion
and contraction of two subspaces during the early asymptotic period into
its converse during the late period.

A ``preferred'' cosmological scenario suggested by these solutions consists
of a combination of the $(-)$ and the $(+)$ branch to account for inflation
as well as for a postinflationary subluminal expansion. 
The apparent shortcoming of this scenario is that the $(-)$ and the
$(+)$ branch constitute two different {\it a priori} unrelated
solutions. As for string frame pre--big--bang models, one might
argue~\cite{pre_BB} that scale factor (T) duality between the branches
provides the correct transition.


\subsection{An Example}

Up to this point our discussion has been rather general. Let us now
illustrate the steps in our solution by giving a simple
example.  
We consider the following situation~: 10-dimensional spacetime is
split into two subspaces with $\bd = (d_0,d_1,0)=(3,6,0)$ and an
elementary IIA RR 3 form occupies the 3--dimensional subspace. This
implies the Ansatz 
\bea
 ds^2 &=& -N^2(\t )d\t^2 +e^{2\a_0}d\bx_0^2+e^{2\a_1}d\bx_1^2\nn \\
 F_{0\m_1\m_2\m_3} &=& e^{-6\a_0}f'(\t )\e_{\m_1\m_2\m_3} \\
 \f &=& \f (\t ) \nn
\eea
in accordance with the eqs.~\refs{metric}, \refs{elementary}.
The equations of motion for this example can be derived from the
Lagrangian
\be
 {\cal L} = E\left[-6{\a_0'}^2-30{\a_1'}^2-36\a_0'\a_1'+\frac{1}{2}
            {\f '}^2+V_e\right] \label{lag_ex}
\ee
with the elementary potential $V_e$ and $E$ given by
\be
 V_e = \frac{1}{2}e^{-6\a_0+\frac{1}{2}\f}{f'}^2\; ,\quad \label{V_ex}
 E = \frac{1}{N}e^{3\a_0+6\a_1} \; .
\ee
which come from the general equations~\refs{lagrangian} and
\refs{Ves}. The equation of motion for $f$ can be integrated to
give the first integral 
\be
 f' = uE^{-1}e^{6\a_0+\frac{1}{2}\f} \label{f_ex}
\ee
with an integration constant $u$. From the Lagrangian~\refs{lag_ex} we
can compute the equations of motion for $\a_0$, $\a_1$ and $\f$. Using
eq.~\refs{f_ex} to replace $f'$ in these equations, we arrive at
\bea
 \frac{d}{d\t}\left( E(-12\a_0'-36\a_1')\right) +3u^2E^{-1}e^{6\a_0-\f /2}
  &=& 0 \nn \\
 \frac{d}{d\t}\left( E(-36\a_0'-60\a_1')\right) &=& 0 \label{eom_ex} \\
 \frac{d}{d\t}\left( E\f '\right)-\frac{1}{4}u^2E^{-1}e^{6\a_0-\f /2}
   &=& 0\nn \; .
\eea
Let us compare these equations with the general ones given in the
moduli space formalism in \refs{al_eom},
\refs{N_eom} and \refs{U}. We see that they can be indeed written in this
form if we set
\be
 G = \left(\ba{ccc} -12&-36&0\\
                      -36&-60&0\\
                       0&0&1\ea\right)\; . \label{G_ex2}
\ee
and
\be
 U=\frac{1}{2}u^2e^{6\a_0-\frac{1}{2}\f} \; . \label{U_ex}
\ee
The matrix $G$ above is consistent with the general formula~\refs{G}
with $d_0=3$, $d_1=6$ and $D=10$. In eq.~\refs{U} we 
introduced a systematic way of writing the effective potential by
introducing a characteristic vector $\bq_r$ for each form. From
eq.~\refs{U_ex} we read off that this vector is given by
$\bq = (6,0,-1/2)$ for our example. This coincides with what one gets by
applying the general rule~\refs{elementary} to the breakup
$(d_0,d_1)=(3,6)$ and a $\d =3$ form in the 3--dimensional subspace. The
dilaton coupling $a(\d )$ for a RR 3--form needed in eq.~\refs{elementary}
follows from eq.~\refs{p_rr} to be $a(\d )=-1/2$.

In section 2 we also defined a scalar product~\refs{s_prod} on the space
of vectors $\bq$ using the inverse of $G$. From eq.~\refs{G_ex2} $G^{-1}$
is given by
\be
 G^{-1} = \left(\ba{ccc} \frac{5}{48}&-\frac{1}{16}&0\\
                         -\frac{1}{16}&\frac{1}{48}&0\\
                         0&0&1\ea\right) \; . \label{Gin_concrete}
\ee
which agrees with the general formula~\refs{Gin} for $(d_0,d_1)=(3,6)$ and
$D=10$. One can easily verify that $<\bq ,\bq > =\bq G^{-1}\bq = 4$.
Therefore $\bq$ is indeed a spacelike vector,
in agreement with the general result~\refs{qq} which showed that this is true
for all vectors obtained from type II forms.

\vspace{0.4cm}

The main problem in solving the system of equations~\refs{eom_ex} is the
existence of two different exponentials, one coming from $E$,
eq.~\refs{V_ex}, the other coming from the effective potential $U$,
eq.~\refs{U_ex}. Fortunately, we have a gauge freedom (time
reparameterization invariance) encoded in $N$ which we can use to get
rid of one of the exponentials. Here, we choose the possibility
of gauging away the potential by setting $E=\exp (6\a_0-\f /2)$.
Given the definition of $E$ in eq.~\refs{V_ex}, this implies
\be
 N=\exp (-3\a_0+6\a_1+\f /2) \label{N_ex}
\ee
in accordance with the general formula~\refs{gauge} for $\bd =(3,6,0)$ and
$\bq = (6,0,-1/2)$. With this choice, the equations of motion~\refs{eom_ex}
turn into
\bea
 \frac{d}{d\t}\left( e^{6\a_0-\f /2}(2\a_0'+6\a_1')\right) &=& \frac{u^2}{2}
  \nn \\
 \frac{d}{d\t}\left( e^{6\a_0-\f /2}(3\a_0'+5\a_1')\right) &=& 0 \\ 
 \frac{d}{d\t}\left( e^{6\a_0-\f /2}\f '\right) &=& \frac{u^2}{2} \nn \\
 e^{6\a_0-\f /2}(6{\a_0'}^2+36\a_0'\a_1'+30{\a_1'}^2-{\f '}^2) &=&
 \frac{u^2}{2} \nn \; .
\eea
This is consistent with the general form~\refs{sol1} found for models with one
term in the potential. Taking appropriate linear combinations of the
first three equations we can derive an equation for the remaining
exponent $6\a_0-\f /2$, which can be solved. Then $\a_0,\a_1,\f$ can be
expressed in terms of this solution. In this way one arrives at
a general solution of the form~\refs{sol1} with coefficients $\bc$
given by
\be
 \bc = \left(\frac{5}{16},-\frac{3}{16},-\frac{1}{4}\right)\; .\label{c_ex2}
\ee
and the following constraints on the integration constants
\bea
 6w_0-\frac{1}{2}w_2 &=& 1 \nn \\
 12w_0^2+72w_0w_1+60w_1^2 &=& w_2^2 \label{cons1_ex} \\
 6k_0-\frac{1}{2}k_2 &=& \ln (u^2) \; .\nn
\eea 
Recall that the time parameter $\t$ is restricted
by $0<\t <\t_1$.

To discuss the cosmology of these solutions we must perform a
transformation to comoving time $t$. To do this, we need the explicit form of
the gauge parameter $N$ which we find by inserting the solution~\refs{sol1}
with~\refs{c_ex2}, \refs{cons1_ex} into eq.~\refs{N_ex}
\be
 N=e^{-3k_0+6k_1+k_2/2}|\t_1-\t |^{-x+\D -1}|\t |^{x-1} \label{Nsol_ex}\; .
\ee
Here $x=3w_0+6w_1$ and $\D =-3/16$. The quantities $x$, $\D$ have been
generally defined in eq.~\refs{x}, \refs{Delta} and their values can be easily
reproduced by inserting $\bd = (3,6,0)$ and $\bc$ from eq.~\refs{c_ex2}.
The range of comoving time obtained by integrating $dt=N(\t )d\t$ over
$0<\t <\t_1$ crucially depends on the singularities in $N$.
Eq.~\refs{Nsol_ex} shows that there are potential singularities at $\t =0$
and $\t =\t_1$. Their appearance is controlled by the value of $x$.
\\[-0.5cm]
\centerline{\epsfbox{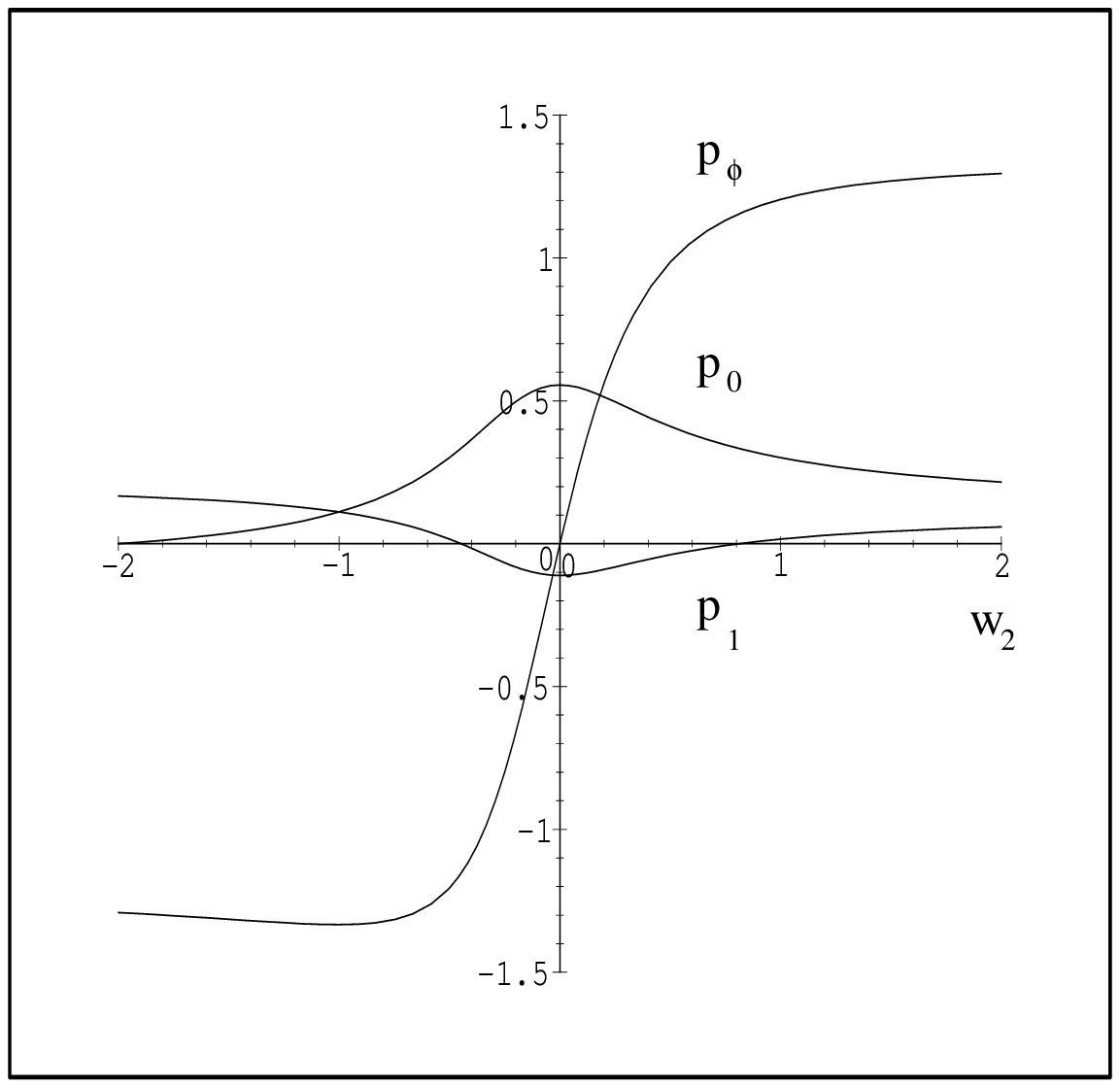}}
\centerline{\em Fig 1: Expansion coefficients for the
            ($+$) branch at $t\simeq t_0$.}
\vskip 0.4cm
Let us therefore analyze which values of $x$ are allowed. The first two
constraints~\refs{cons1_ex} may be solved to express, for example, $w_0$
and $w_1$ as a function of $w_2$. This shows that $x$ depends on one
free parameter only. Furthermore, since the second constraint~\refs{cons1_ex}
is quadratic in $w_I$, we find two branches satisfying $x<\D = -3/16$ and
$x>0$, respectively. We refer to these two branches as the $(-)$ and
the $(+)$ branch. From eq.~\refs{Nsol_ex} we see that $0<\t <\t_1$ is
indeed mapped to the comoving time ranges given in eq.~\refs{ranges};
that is to $\left[ -\infty ,t_1\right]$ for the $(-)$ branch and to
$\left[ t_0,\infty\right]$ for the $(+)$ branch ($t_0,t_1$ are integration
constants). Moreover, the scalar curvature~\refs{R} has a future timelike
singularity in the $(-)$ branch and a past timelike singularity in the
$(-)$ branch. Both types of solutions are therefore not extendible.\\
\centerline{\epsfbox{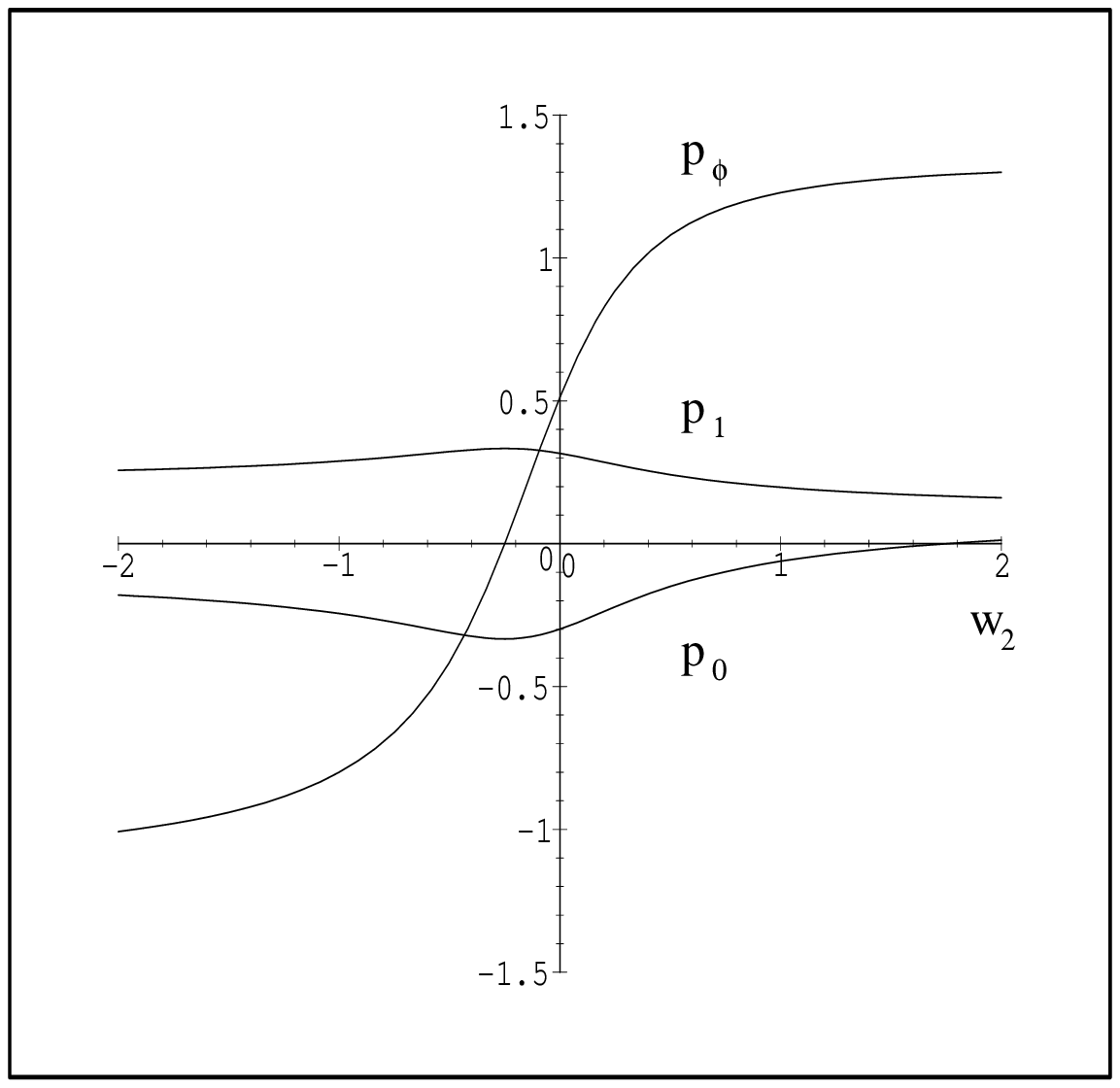}}
\centerline{\em Fig 2: Expansion coefficients for the ($+$) branch at
            $t\rightarrow\infty$.}
\vskip 0.4cm
Information about the evolution of the two subspaces and the dilaton
can be obtained form the respective Hubble parameters $\bH =\dot{\bal}$
written as a function of comoving time. They can be calculated if
$\t\simeq 0$ or $\t\simeq\t_1$ since $N$ in eq.~\refs{Nsol_ex} can be
integrated in these limits. Doing this for our example by using
eq.~\refs{sol1}, \refs{c_ex2}, \refs{cons1_ex} and $N$, $\t (t)$
calculated from eq.~\refs{Nsol_ex}, we find $\bH$ to be of the
Kaluza--Klein form~\refs{hubble}, \refs{kk}. The expansion coefficients
$\bp$ depend on the integration constants $\bw$ as in eq.~\refs{p_expr}.

In fact, using the first two constraints~\refs{cons1_ex} we may rewrite
$\bp$ as a function of $w_2$ only, as we did for the parameter $x$ before.
The asymptotic expansion properties of our example therefore depend on
one free parameter only. Instead of giving the explicit formulae,
which are not particularly enlightening, let us give a graphical 
representation of $\bp =\bp (w_2)$. We concentrate on the $(+)$ branch (the
expansion coefficients in the $(-)$ branch can be worked out analogously)
where $t_0<t<\infty$ and the asymptotic regions are characterized by
$t\simeq t_0$ ($\t\simeq 0$) and $t\rightarrow\infty$ ($\t\simeq\t_1$). 
The results are given in fig.~1 ($t\simeq t_0$) and fig.~2
($t\rightarrow\infty$).
In both figures $|p_0|,|p_1|<1$ always, which illustrates our general
result that the expansion in the $(+)$ branch is always subluminal.
We see that an early expansion of the 3--dimensional subspace
($p_0>0$ in fig.~1) is turned into a contraction as $t\rightarrow\infty$
($p_0<0$ in fig.~2) for a wide range in $w_2$. This can be understood from
the $\a_0$ dependence of the effective potential~\refs{U_ex}. Moreover, the
6--dimensional space is always expanding as $t\rightarrow\infty$
($p_1>0$ in fig.~2). In a more realistic model, such an expansion should be
stopped by, for instance, a nonperturbative potential for the modulus $\a_1$.

\section{Ricci Flat Compactifications of 11--Dimensional Supergravity}

So far, we have considered cosmological M--theory solutions which
correspond to products of maximally symmetric spaces. We would now like
to shift our attention to a different, partially overlapping class of
configurations, namely those with internal Ricci--flat spaces. In particular,
this allows for more realistic models with the internal space being a
Calabi--Yau manifold. We will find that the characteristic features of
such solutions agree with the ones we discovered previously for the space
being a product of maximally symmetric subspaces. Especially, we confirm
the r\^ole of the form field as mapping different Kaluza--Klein type
expansions into each other. In addition, we study an important aspect
of M--theory which we have not addressed earlier, namely U--duality
symmetries and their relation to cosmological solutions. For a related
approach to cosmology using T--duality see~\cite{Tcosm,mv}.

Let us start by reducing the bosonic part of 11--dimensional
supergravity to $D$ space--time dimensions on a Ricci--flat manifold.
The resulting low energy action will be the starting point for our
discussion of cosmological solutions.

The bosonic part of the 11--dimensional supergravity Lagrangian reads
\be
 \cL = \sqrt{-\bar{g}}\left[ \bar{R}-\frac{1}{2\cdot 4!} F_{MNPQ}
       F^{MNPQ}\right] +\frac{1}{3\cdot 3!(4!)^2}
       \,\e^{M_1...M_{11}}{F}_{M_1..M_4}F_{M_5..M_8}
       A_{M_9M_{10}M_{11}}\; . \label{SUGRA11}
\ee
We are using the conventions of ref.~\cite{cj}. The 11--dimensional metric
and curvature are given by $\bar{g}_{MN}$ and $\bar{R}$, respectively, where
uppercase letters are used to index the full space, that is,
$M,N,...=0,...,10$. The 4--form field strength $F_{MNPQ}$ is expressed
in terms of the 3--form gauge field $A_{NPQ}$ as
$F_{MNPQ}=4\,\partial_{[M}A_{NPQ]}$. For the class of
compactifications we will be interested in, the Chern--Simons term in
eq.~\refs{SUGRA11} vanishes. Therefore, we drop this term from now on and
consider the non--topological part of the Lagrangian
\be
 \cL = \sqrt{-\bar{g}}\left[ \bar{R}-\frac{1}{2\cdot 4!}F_{MNPQ}
       F^{MNPQ}\right] \label{L}
\ee
only. Our main purpose is to investigate the relation of cosmological
solutions and U--duality symmetries for the action~\refs{L}. The focus in this
paper will be on the external part of U--duality which acts non--trivially
on the space--time metric rather than on the part which transforms moduli.
In our reduction to $D$ dimensions we will therefore keep a minimal moduli
content only, that is, the breathing mode of the Ricci--flat manifold.
Though formulae in this section are kept general, the most interesting cases
are the ones for $D=4,5$. While the case
$D=4$ is of obvious relevance, the importance of $D=5$ is motivated by
the construction of Horava and Witten~\cite{hw} for the effective action
of the strongly coupled heterotic string. 

\vspace{0.4cm}

Let us now be specific. We are using indices $\m ,\n ...=0,...,d\equiv
D-1$ for the external space--time, indices $m,n,...=1,...,d$ for the
external spatial directions and indices $a,b,...=d+1,...,10$ for the
$\d$--dimensional internal space, where $\d = 11-D$. Our Ansatz for the
11--dimensional fields is as follows
\bea
 \bar{g}_{\m\n} &=& \bar{g}_{\m\n}(x^\r ) \nn \\
 \bar{g}_{\m b} &=& 0 \nn \\
 \bar{g}_{ab} &=& \bar{b}^2(x^\r )\O_{ab}(x^c) \label{ans} \\
 A_{\m\n\r} &=& B_{\m\n\r} (x^\s )\; . \nn
\eea
All other components of $A_{NPQ}$ are set to zero. Here  
$\O_{ab}$ is the metric of a $\d$--dimensional Ricci--flat manifold and
$\bar{b}$ is its breathing mode. Depending on the dimension, this manifold
can be a Calabi--Yau space, a torus or even a product of both.
The $D$--dimensional metric and 3--form are denoted
by $\bar{g}_{\m\n}$ and $B_{\m\n\r}$, respectively. As already discussed,
we have considered the minimal moduli content represented by $\bar{b}$ only.
As a further simplification, we have neglected $D$--dimensional
vector fields (graviphotons as well as those arising from $A_{NPQ}$)
and $D$--dimensional 2 forms. With the above truncation, we arrive at a
low energy theory independent on the details of the compactification,
but we keep the characteristic 3--form as a low energy field. Furthermore,
as we will see, the Ansatz~\refs{ans} is consistent with the external part of
U--duality, so that it provides a ``minimal'' setting for our purpose.

Using the truncation~\refs{ans} the action~\refs{L} turns into
\be
 \cL = \sqrt{\O}\,\sqrt{-\bar{g}}\;\bar{b}^\d\left[ \bar{R}+
       \d (\d -1)\,\bar{b}^{-2}(\partial_\m\bar{b})^2-\frac{1}{2\cdot 4!}
       F_{\m\n\r\s}F^{\m\n\r\s}\right]\; , \label{L1}
\ee
where $F_{\m\n\r\s}=4\,\partial_{[\m}B_{\n\r\s ]}$. To get a canonical
curvature term we perform the Weyl rotation
\be
 \bar{g}_{\m\n} = \bar{b}^{-\frac{2\d}{D-2}}\,g_{\m\n} \label{Weyl}
\ee
to the Einstein frame metric $g_{\m\n}$. In this frame, eq.~\refs{L1}
reads~\footnote{We drop the factor $\sqrt{\O}$ since it turns into a 
constant upon integration over the internal manifold.}
\be
 \cL = \sqrt{-g}\left[ R-k^2\,\bar{b}^{-2}(\partial_\m\bar{b})^2
       -\bar{b}^{\frac{6(11-D)}{D-2}}\frac{1}{2\cdot 4!}F_{\m\n\r\s}
       F^{\m\n\r\s}\right] \label{L2}
\ee
with $k^2=\frac{D-1}{D-2}\d^2-\d (\d -1)$. For a study of cosmological
solutions of this Lagrangian we consider the Ansatz
\bea
 g_{\m\n} &=& \left(\ba{cc} -\bar{N}^2(\t ) & 0 \\
                                    0 & \bar{G}_{mn}(\t )\ea\right) \nn \\
 B_{mnr} &=& B_{mnr}(\t ) \\
 \bar{b} &=& \bar{b}(\t ) \nn \; .
\eea
Here, time has been denoted by $\t$.
The equations of motion with these specialized fields inserted can be
derived from a Lagrangian which is related to eq.~\refs{L2} by a formal
dimensional reduction to one dimension. This 1--dimensional Lagrangian
is given by
\bea
 \cL &=& \bar{N}^{-1}\sqrt{\bar{\F}}\left[ k^2\,\bar{b}^{-2}{\dot{\bar{b}}}^2
       -\frac{1}{4}{\bar{\F}}^2 {\dot{\bar{\F}}}^2-
       \frac{1}{4}\dot{\bar{G}}_{mn}{\dot{\bar{G}}}^{mn}\right. \nn \\
      && \quad\quad\quad\quad\quad
         \left. +\bar{b}^{\frac{6(11-d)}{D-2}}\bar{G}^{mm'}\bar{G}^{nn'}
         \bar{G}^{rr'}\dot{B}_{mnr}\dot{B}_{m'n'r'}\right] \label{L3}
\eea
where $\bar{\F}={\rm det}(\bar{G})$. Unlike in the first sections of this
paper, here
the dot denotes the derivative with
respect to the time $\t$. 

In the last step we have dimensionally reduced $d=D-1$ dimensions of a theory
which by itself has been obtained reducing 11--dimensional
supergravity. Therefore, one should expect the U--duality group of
$(11-d)$--dimensional maximal supergravity as a symmetry group of the
Lagrangian~\refs{L3}. For example, for $D=4$ ($d=3$) the expected group is
the one of $8$--dimensional supergravity, that is, $G=SL(2)\times SL(3)$.
For $D=5$ ($d=4$) one expects $G=SL(5)$, the U--duality group of
7--dimensional supergravity. As we will show, this is indeed the case.
It is, however, hard to see directly from the Lagrangian in the form~\refs{L3}.
The reason is that we have performed a Weyl rotation~\refs{Weyl}
which is different from the one that leads to $(11-d)$--dimensional
supergravity with a canonical Einstein term. We can compensate for this
by the following nonlinear field redefinitions
\bea
 \bar{b} &=& \F^{-\frac{1}{2(10-D)}}\, b \nn\\
 \bar{N} &=& \F^{-\frac{9}{2(10-D)(D-2)}}\, b^{\frac{11-d}{D-2}}\, N \nn \\
 \bar{G}_{mn} &=& \F^{-\frac{11-d}{(10-D)(D-2)}}\,
                  b^{\frac{2(11-D)}{D-2}}\, G_{mn} \label{trafo}\\
 \bar{\F} &=& \F^{-\frac{(11-D)(D-1)}{(10-D)(D-2)}+1}\, b^{\frac{2(11-D)(D-1)}
              {D-2}} \nn\; ,
\eea
which express the physical Einstein frame fields $\bar{G}_{mn}$,$\bar{N}$,
$\bar{b}$, $\bar{\F}$ in terms of the new fields $G_{mn}$, $N$, $b$, $\F$
with 
\be
 \F = {\rm det}(G)\; .
\ee
Written in terms of these new variables the Lagrangian~\refs{L3} finally
reads
\bea
 \cL &=& N^{-1}\, b^\d\left[ -\d (\d -1)\,b^{-2}{\dot{b}}^2+\frac{1}{4(10-D)}
       \F^{-2}{\dot{\F}}^2-\frac{1}{4}\dot{G}_{mn}{\dot{G}}^{mn}\right.\nn\\
      &&\quad\quad\quad\quad
        \left. +\frac{1}{2\cdot 3!}G^{mm'}G^{nn'}G^{rr'}\dot{B}_{mnr}
        \dot{B}_{m'n'r'}\right]\; . \label{Lf}
\eea
This is the form of $\cL$ we are going to use in our discussion of U--duality
and cosmological solutions. For a physical interpretation of solutions one
should, of course, transform back to the Einstein frame fields
via eq.~\refs{trafo}.

\section{U--Duality Covariant Formulation}

In this section, we will find the manifestly U--duality invariant form of
the Lagrangian~\refs{Lf} and discuss its general solution.

\vspace{0.4cm}

The cases $D=4,5$ can be treated uniformly by considering an
$SL(n)/SO(n)$ sigma model (where $n=2,3$ for $D=4$ and $n=5$ for $D=5$, for
example)
written in terms of the coset parameterization $M\in SL(n)/SO(n)$. 

Without reference to a specific parameterization, the coset $M$ can be
characterized by the conditions ${\rm det}(M)=1$ and $M=M^T$ which
can be implemented via Lagrange multipliers. We are therefore considering
the Lagrangian
\be
 \cL_1 = N^{-1}\, b^\d\left[ -\d (\d -1)\, b^{-2}{\dot{b}}^2+\frac{1}{4}
         {\rm tr}\left( M^{-1}\dot{M}M^{-1}\dot{M}\right)\right]
         +\l \left({\rm det}(M)-1\right)+{\rm tr}\left(\g (M-M^T)\right)
 \label{Lm}
\ee
with the Lagrange multipliers $\l$, $\g$. The $SL(n)$ symmetry transformations
are given by
\bea
 b \rightarrow b&,& \l \rightarrow \l \nn\\
 M \rightarrow PMP^T &,& \g \rightarrow {P^T}^{-1}\g P^{-1}\; ,
 \label{sln_trafo}
\eea
where $P\in SL(n)$. 
After eliminating the Lagrange multipliers, we find as the $SL(n)$
covariant equations of motion for $M$, $b$ and $N$ (in the gauge $N=1$ which
we can always choose by a suitable reparameterization of the time $\t$)
\bea
 \frac{d}{d\t}\left( M^{-1}\dot{M}\right) +\d HM^{-1}\dot{M} &=& 0 \nn \\
 (\d -1)\dot{H}+\frac{1}{2}\d (\d -1)H^2 &=& -\r \label{eoms}\\
 \frac{1}{2}\d (\d -1)H^2 &=& \r\; ,\nn
\eea
respectively. Clearly, the matrix $M$ in these equations is restricted to
be symmetric and unimodular. The Hubble constant $H$ and the energy density
$\r$ are defined by
\be
 H = \frac{\dot{b}}{b}\; ,\quad\quad \r = \frac{1}{8}{\rm tr}\left(
     M^{-1}\dot{M}M^{-1}\dot{M}\right)\; .\label{Hrho}
\ee
The second and third equation in~\refs{eoms} can be combined to find the
following solution for the breathing mode 
\be
 H = \frac{1}{\d \t}\; ,\quad\quad b=b_0\, |\t |^{1/\d}\; ,\label{b_sol}
\ee
where $b_0$ is an arbitrary constant. Inserting this into the first
equation~\refs{eoms} we find for the coset $M$
\be
 M = M_0\, e^{I\ln |\t |}\; ,\label{sol_M}
\ee
where the constant matrices $M_0$, $I$ satisfy
\bea
 {\rm det}(M_0) = 1&,&\quad {\rm tr}(I) = 0 \nn \\
 M_0=M_0^T &,&\quad M_0I=I^TM_0\; . \label{sol_cons}
\eea
Furthermore, from the last equation~\refs{eoms} one obtains the zero energy
constraint
\be
 {\rm tr}(I^2) = 4\frac{\d -1}{\d}\; .\label{E0}
\ee
Eqs.~\refs{b_sol}--\refs{E0} represent the complete solution of the
Lagrangian~\refs{Lm} written in a manifestly $SL(n)$ covariant form.
The $SL(n)$ transformation~\refs{sln_trafo} on the coset $M$ acts on the
integration constants encoded in $M_0$, $I$ by
\bea
 M_0&\rightarrow& PM_0P^T \nn \\
 I&\rightarrow& {P^T}^{-1}IP^T\; .\label{sln_par}
\eea
At this point, it is instructive to count the number of integration constants
in our general solution. The matrices $M_0$, $I$ satisfying the
constraints~\refs{sol_cons} contain $n^2+n-2$ independent parameters. The
zero energy condition~\refs{E0} eliminates one of them so that we remain with
$n^2+n-3$ independent integration constants. This is just about the correct
number to describe the general solution for all degrees of freedom in
the coset $M\in SL(n)/SO(n)$. On the other hand, the group $SL(n)$
consists of $n^2-1$ parameters which implies that for $n>2$ not all solutions
can be connected to each other by $SL(n)$ transformations. More precisely,
the total $n^2+n-3$--dimensional solution space splits into
$n^2-1$--dimensional equivalence classes, each consisting of solutions
related to each other by $SL(n)$ transformations via eq.~\refs{sln_par}.
The remaining $n-2$ integration constants label different equivalence classes,
that is, classes of solutions which cannot be connected by $SL(n)$
transformations. It is useful in the following to make this structure more
explicit in the solution~\refs{sol_M}. Diagonalizing $M_0$ and $I$ using
eq.~\refs{sln_par} with appropriate matrices $P$, it is straightforward to
prove that eqs.~\refs{sol_M}, \refs{sol_cons}, \refs{E0} can equivalently
be written in the form
\be
 M = P\, {\rm diag}(|\t |^{p_1},...,|\t |^{p_n})\, P^T
 \label{sol_M1}
\ee
with
\be
 \sum_{i=1}^{n}p_i = 0\; ,\quad\quad \sum_{i=1}^{n}p_i^2=4\frac{\d -1}{\d}
 \label{p_cons}
\ee
and $P\in SL(n)$. The equivalence classes of $SL(n)$ unrelated solutions
are parameterized by the $n-2$ constants $\{ p_i\}$ subject to the
constrains~\refs{p_cons}. In addition, since $SL(n)$ contains
permutations of the $n$ directions we should pick a definite order, say
$p_i\geq p_j$ if $i<j$, for the $\{ p_i \}$ to describe $SL(n)$ inequivalent
solution. On the other hand, a specific class, characterized
by a fixed set $\{ p_i\}$, is generated by the matrices $P\in SL(n)$ in
eq.~\refs{sol_M1}. In the next section, we will apply these general results
to the example $D=5$ and discuss the physical implications.

\section{The Example $D=5$}

The U--duality group in the $D=5$ case is $G=SL(5)$. Let us
define the vector ${\bf B} = (B^s)$ by
\be
 B_{mnr} = \frac{1}{\F}\e_{mnrs}B^s\; .
\ee
Then the $SL(5)/SO(5)$ coset $\cM$ can be parameterized by~\cite{dl}
\be
 \cM = \F^{-2/5}\left(\ba{cc} G & -G{\bf B} \\
                              -{\bf B}^TG & \F+{\bf B}^TG{\bf B}\ea\right)\; ,
 \label{cos5}
\ee
where we have used a matrix notation $G=(G_{mn})$ for the metric. With
the internal dimension $\d =6$, Lagrangian~\refs{Lf} can then be put into
the form
\be
 \cL = N^{-1}b^6\left[ -30\, b^{-2}{\dot{b}}^2-\frac{1}{4}{\rm tr}\left(
       \dot{\cM}{\dot{\cM}}^{-1}\right)\right]\; ,
 \label{L5}
\ee
which has manifest $SL(5)$ invariance. The explicit transformations are
given by
\bea
 b &\rightarrow& b \nn \\
 \cM &\rightarrow& P\cM P^T
\eea
with $P\in SL(5)$.

We are now dealing with a $\d =6$--dimensional internal manifold which
can be a torus $T^6$ or a Calabi--Yau 3--fold. The equations of motion
for the Lagrangian~\refs{L5} are given by the general
expressions~\refs{eoms}, \refs{Hrho} with $\d =6$ and $M=\cM$ inserted.
Here $\cM$ is the $SL(5)/SO(5)$ coset explicitly given in terms of the
metric and the 3--form in eq.~\refs{cos5}. From eq.~\refs{trafo} the
physical fields can be written as
\bea
 \bar{b} &=& \F^{-1/10}\, b \nn \\
 \bar{N} &=& \F^{-3/10}\, b^2 \label{trafo5}\\
 \bar{G}_{mn} &=& \F^{-2/5}\, b^4\, G_{mn}\nn\; .
\eea
The solution for the breathing mode can be read off from eq.~\refs{b_sol}
\be
 H=\frac{1}{6\t}\; ,\quad b=b_0\, |\t |^{1/6}\; .
\ee
For the coset $\cM$ we have from eq.~\refs{sol_M1} and \refs{p_cons}
\be
 \cM = P\, {\rm diag}(|\t |^{p_1},...,|\t |^{p_5})P^T\; ,\label{sol_M5}
\ee
with
\be
 \sum_{i=1}^5p_i=0\; ,\quad \sum_{i=1}^5p_i^2=\frac{10}{3}\label{p_cons5}
\ee
and $P\in SL(5)$. As before, we require $p_i\geq p_j$ for $i<j$. The
solution~\refs{sol_M5} contains 27 integration
constants. Three of them are given by the parameters $\{ p_i\}$ subject
to the constraints~\refs{p_cons5}, labeling the $SL(5)$ equivalence classes.
The remaining 24 integration constants parameterize the $SL(5)$ matrix
$P$ in eq.~\refs{sol_M5}.

What is the general physical picture emerging from the solution~\refs{sol_M5}?
As we will see explicitly below, our solutions have two different branches
which correspond to the two different signs of $\t$ in eq.~\refs{sol_M5}.
These branches are the $(-)$ and $(+)$ branches we have found earlier.
From the structure of eqs.~\refs{sol_M5} and \refs{trafo5} it is clear
that, depending on the choice of the $SL(5)$ matrix $P$, physically relevant
quantities like Hubble expansion parameters will in general depend on
linear combinations of the various exponents $|\t |^{p_i}$. In certain
asymptotic regions, however, one of these exponents will usually dominate
and the expansion is described by simple power laws. These asymptotic regions
correspond to the Kaluza--Klein regions found previously. The interesting
observation here is, that the Hubble expansion rates in those Kaluza--Klein
regions are U--duality invariant since this is the case for the powers $p_i$.
A related observation for S--duality transformations has been made in
ref.~\cite{cew1,clw0,clw}. The choice of the $SL(5)$ matrix $P$, on the
other hand, determines the time range for the asymptotic regions and the
details of the transition.

Though generically clear, this picture is rather complicate to analyze in
detail for the general solution~\refs{sol_M5}. Therefore we concentrate
on the physically interesting case of FRW universes in the following. 
Consequently, we require a 3--dimensional spatial subspace of our
5--dimensional space to be isotropic. For the metric and the form field
this implies
\be
 G=\left(\ba{cc} c{\bf 1}_3&0\\0&\f\ea\right)\; ,\quad 
 {\bf B} = \left(\ba{c}{\bf 0}_3\\B\ea\right)\; ,\label{frw5}
\ee
where $c,\f ,B$ are time dependent scalars. Inserting~\refs{frw5} into
the coset parameterization~\refs{cos5} for $\cM$ results in
\be
 \cM =\left(\ba{cc}\cM_3&0\\0&\cM_2\ea\right)\; ,\quad
 \cM_3 = c\F^{-2/5}{\bf 1}_3\; ,\quad \cM_2 = \F^{-2/5}\left(\ba{cc}
         \f&-\f B\\ -\f B&\F +B^2\f\ea\right)\; ,\label{cos_frw}
\ee
with $\F =c^3\f$. Eq.~\refs{cos_frw} shows that (unlike in the
case $D=4$, see ref.~\cite{ucos}) the property ``FRW universe'' is not
invariant under the full U--duality group $SL(5)$. In particular, FRW
universes can be mapped into anisotropic solutions and vice versa using
appropriate $SL(5)$ transformations. Since we wish to stay within the class
of FRW solutions, we should restrict ourselves to the subgroup
$H\equiv SL(3)\times SL(2)\times U(1)\subset SL(5)$ which leaves the structure
of $\cM$ in eq.~\refs{cos_frw} invariant. Explicitly, this subgroup acts
as
\be
 \cM_{2,3}\rightarrow P_{2,3}\cM_{2,3}P_{2,3}^T\label{Hact}
\ee
with $P_{2}\in GL(2)$, $P_{3}\in GL(3)$ and ${\rm det}(P_2){\rm det}(P_3)=1$.
Let us consider equivalence classes of solutions with respect to this
subgroup $H$ instead of the full group $SL(5)$. Then FRW universes are
specified by
\be
 p\equiv p_1=p_2=p_3
\ee
in eq.~\refs{sol_M5}. From eq.~\refs{p_cons5} we derive
\be
 p_{4,5} = -\frac{3p}{2}\pm\sqrt{\frac{5}{3}-\frac{15}{4}p^2}\; ,\quad
 |p|\leq \frac{2}{3}\; .
 \label{p45}
\ee
We have therefore found a one parameter set (with parameter $p$) of
$H$--inequivalent classes of FRW universes, each equivalence class for a
fixed $p$ spanned by the action of the group $H$ in eq.~\refs{Hact}.
How does $H$ act explicitly? First of all, $GL(3)\subset H$ is again
part of the global coordinate transformations and therefore trivial.
We concentrate on the $SL(2)$ part and write
\be
 P_2 = \left(\ba{cc} \a&\b\\ \g&\d\ea\right)\; ,\quad \a\d -\b\g = 1\; .
\ee
Then, $\cM_2$, $\cM_3$ take the form
\be
 \cM_2 = \left(\ba{cc}\a^2|\t |^{p_4}+\b^2|\t |^{p_5}&
                      \a\g |\t |^{p_4}+\b\d |\t |^{p_5}\\
                      \a\g |\t |^{p_4}+\b\d |\t |^{p_5}&
                      \g^2|\t |^{p_4}+\d^2|\t |^{p_5}\ea\right)\; ,
 \quad\cM_3 = |\t |^p{\bf 1}_3\; ,
\ee
where $p_{4,5}$ are given by~\refs{p45} in terms of the free parameter $p$.
By comparison with eq.~\refs{cos_frw} we can read off the expressions
for $c,\F ,B, \f$ and convert them to the physical fields via
eq.~\refs{trafo5}. The result is
\bea
 \bar{b} &=& b_0\, (\a^2|\t |^{p_4}+\b^2|\t |^{p_5})^{1/6}|\t |^{p/2+1/6}
            \nn \\
 \bar{N} &=& b_0^2\, (\a^2|\t |^{p_4}+\b^2|\t |^{p_5})^{1/2}|\t |^{3p/2+1/3}
  \nn \\
 \bar{G}_{mn} &=&a^2\,\d_{mn}\; ,\quad a = b_0^2\, (\a^2|\t |^{p_4}
                 +\b^2|\t |^{p_5})^{1/3}|\t |^{p+1/3} \\
 B &=& -\frac{\a\g |\t |^{p_4}+\b\d |\t |^{p_5}}
       {\a^2|\t |^{p_4}+\b^2|\t |^{p_5}}\; .\nn
\eea
We can solve for the comoving time $t$ in two asymptotic regions leading to
\bea
 t &=& \frac{2b_0^2\, |\a |}{3p+p_4+8/3}|\t |^{3p/2+p_4/2+4/3}\, {\rm sgn}(\t )
       \; , \quad {\rm for}\quad |\t |\gg\t_{\rm form} \nn \\
 t &=& b_0^2\, |\b |\frac{2}{3p+p_5+8/3}|\t |^{3p/2+p_5/2+4/3}\, {\rm sgn}(\t )
       \; , \quad {\rm for}\quad |\t |\ll\t_{\rm form}\; ,
\eea
where
\be
 \t_{\rm form} = \left(\frac{\b}{\a}\right)^{\frac{2}{p_4-p_5}}\; .
\ee
In these regions, we find for the Hubble parameter $H_a$
\be
 H_a = \frac{P(p)}{t}\; ,\quad P(p) = \left\{\ba{lllll}
       \frac{2}{3}\frac{3p+p_4+1}{3p+p_4+8/3}&&{\rm for}
       &&|\t |\gg\t_{\rm form}\\
       \frac{2}{3}\frac{3p+p_5+1}{3p+p_5+8/3}&&{\rm for}
       &&|\t |\ll\t_{\rm form}\ea\right.\; .
       \quad \label{Hub5}
\ee
The expansion coefficient $P(p)$ depends on the free parameter $p$ and is
generically different in the two asymptotic regions. As can be seen from
fig.~1, it is always positive for large $|\t |\gg \t_{\rm form}$
and can have both signs for $|\t |\ll \t_{\rm form}$. For the
positive branch $t>0$ this implies a universe which is expanding or
contracting at early time and is turned into an expanding universe at late
time. The situation for the negative branch is reversed; the universe
is always contracting at early time ($|t|$ large and $t<0$) and can
be contracting or expanding later. As before, the Hubble parameter~\refs{Hub5}
and hence the aforementioned properties are $SL(2)$ invariant. The transition
time given by
\be
 |\t |\sim\t_{\rm form}=\left(\frac{\b}{\a}\right)^{\frac{2}{p_4-p_5}}\; ,
\ee
on the other hand, depends on $SL(2)$ parameters along with the details of
the transition.\\
[-0.5cm]

\epsfbox[-80 0 500 210]{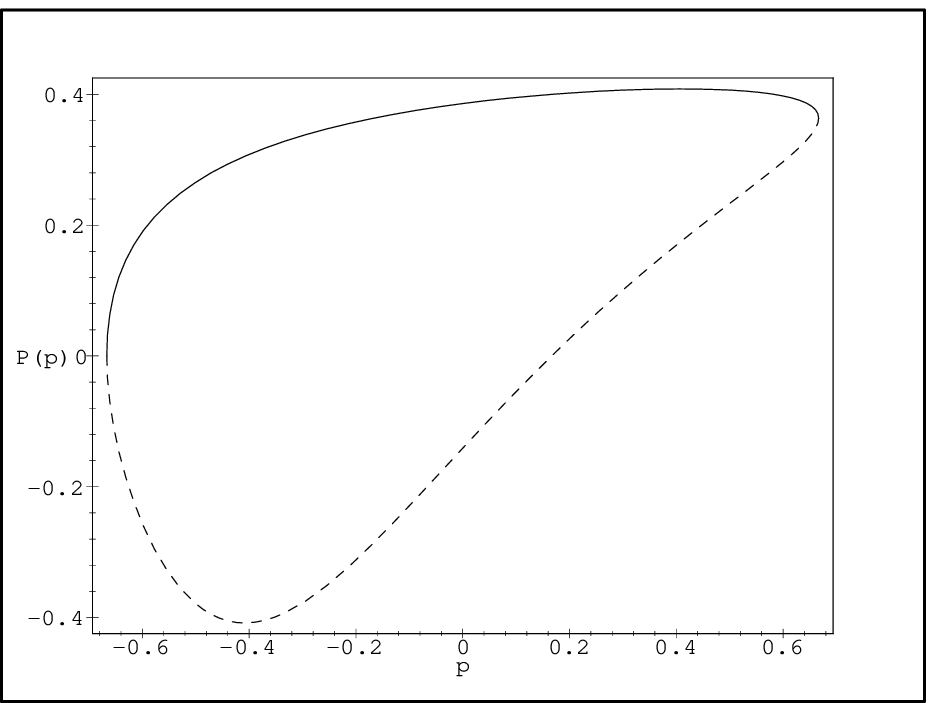}
\centerline{\em Fig 1: Expansion coefficient $P(p)$ for
$|\t |\gg\t_{\rm form}$ (solid curve) and $|\t |\ll\t_{\rm form}$
(dashed curve).}
\vskip 0.4cm

Suppose now, we choose a solution in the positive branch which is contracting
for a short period of time and then turns into expansion. By applying
appropriate $SL(2)$ transformations to this solution the contraction period
can be made arbitrarily long. An additional time reversal $t\rightarrow -t$
leads to a negative branch solution with an expansion period that can be
arbitrarily long. The extreme limits are possible. By choosing $P_2={\bf 1}$
($\b =0$ in particular) we have a positive branch solution which is always
expanding. As the other extreme we may set
\[
 P_2 = \left(\ba{cc}0&1\\-1&0\ea\right)
\]
($\a =0$ in particular) which
generates an expanding negative time branch solution. We have therefore shown
that a combination of U--duality and time reversal can map expanding
negative and positive time branch solutions into each other. The expansion
coefficients in the negative and positive branch then correspond to the
lower and upper part of the curve in fig.~1. An analog mapping, carried
out by a T--duality transformation combined with a time reversal, is the
starting point of the pre--big--bang scenario~\cite{pbb} of weakly coupled
heterotic string cosmology.

\vspace{0.4cm}


\vspace{0.8cm}

{\bf Acknowledgments} A.~L.~is supported by a fellowship from Deutsche
Forschungsgemeinschaft (DFG). A.~L.~and B.~A.~O.~are supported in part by
DOE under contract No. DE-AC02-76-ER-03071. D.~W.~is supported in part by
DOE under contract No. DE-AC02-76-ER-03072. 
\end{document}